\documentstyle[12pt,epsf]{article}

\input{epsf}
\begin{document}
\begin{titlepage}
\begin{center}

{\Large \bf An Exact Approach to the Oscillator Radiation Process in an Arbitrarily Large Cavity}\\
\vspace{.3in}
{\large\em N.P. Andion${}^{(b)}$, A.P.C. Malbouisson${}^{(a)}$ and A. Mattos Neto${}^{(b)}$} \\
\vspace{0.1cm}
${}^{(a)}$ {\it Centro Brasileiro de Pesquisas F\'{\i}sicas,} \\
\vspace{0.1cm}
{\it Rua Dr. Xavier Sigaud 150, Urca,} \\
\vspace{0.1cm}
{\it Rio de Janeiro CEP 22290-180-RJ, Brazil.} \\
\vspace{0.1cm} 
{\it E-mail: adolfo@lafex.cbpf.br} \\
\vspace{0.7cm}
${}^{(b)}$ {\it Instituto de Fisica - Universidade Federal da Bahia
} \\
\vspace{0.1cm}
{\it Campus Universitario de Ondina, 40210-340-BA Salvador Brazil} \\
\vspace{0.1cm} 
{\it E-mail: andion@ufba.br,\,arthur@fis.ufba.br} \\
\vspace{0.7cm}

\end{center}

\subsection*{Abstract}
Starting from a solution of the problem of a mechanical oscillator coupled to a scalar field inside a reflecting sphere of radius $R$, we study the behaviour of
  the system in free space as the limit of an arbitrarily large
 radius in the confined solution. From
a mathematical
point of view we show that this way of facing the problem is not equivalent to consider
the system {\it a} {\it priori} embedded in infinite space. In particular, the 
matrix elements of the transformation turning the system to principal axis, do not tend
to distributions in the limit of an arbitrarily large sphere as it should be the case 
if the two procedures were mathematically equivalent. Also, we introduce "dressed" coordinates
which allow an exact description of the oscillator radiation process for any value of the 
coupling, strong or weak.
In the case of weak coupling, we recover from our exact expressions the well known decay 
formulas from perturbation theory.

\end{titlepage}
\newpage \baselineskip .37in

\section{Introduction}

Since a long time ago 
the experimental and theoretical investigations on the polarization of atoms
by optical pumping and the possibility of detecting changes in their
polarization states has allowed the observation of resonant effects
associated to the coupling of these atoms with strong radiofrequency fields 
\cite{Winter}. As remarked 
in \cite{bouquinCohen}, the theoretical understanding of these effects using
perturbative methods requires the calculation of very high-order terms in
 perturbation theory, what makes the standard Feynman
diagrams technique practically unreliable in those cases. The trials of treating
non-perturbativelly such kind of systems consisting of an atom coupled to the
electromagnetic field, have lead to the idea of "dressed atom", introduced in refs 
\cite{Polonsky} and \cite{Haroche}. This approach consists in quantizing the electromagnetic
field and analyzing the whole system consisting of the atom coupled to the
electromagnetic field.
Along the 
 years since then, this concept has been extensively used  to investigate
several situations involving the interaction of atoms and electromagnetic
fields. For instance, atoms embedded in a strong radiofrequency field
background in refs. \cite{Cohen1} and  \cite
{Cohen2}, atoms in intense resonant laser beans in ref. \cite
{Haroche1} or the study of photon correlations and quantum jumps. In this
last situation, as showed in refs. \cite{Cohen3}, \cite
{Cohen4} and \cite{Cohen5}, the statistical properties of the random sequence
of outcoming pulses can be analyzed by a broadband photodetector and the
dressed atom approach provides a convenient theoretical framework to perform
this analysis.

Besides the idea of dressed atom in itself, another aspect that desserves attention
is the non-linear character of the problem involved in realistic
situations, which implies, as noted above, in very hard mathematical problems to be 
dealt with. An way to circunvect these mathematical difficulties, is to assume that under certain
conditions the coupled atom-electromagnetic field 
system may be approximated by the
system composed of an harmonic oscillator coupled {\it linearly} to the
field trough some effective coupling constant $g$.
   
In this sense, 
in a slightly different context, recently a significative number of works
has been spared to the study of cavity QED, in particular to the theoretical
investigation of higher-generation Schrodinger cat-states in high-Q
cavities, as has been done for instance in \cite{Jmario}. Linear approximations of this type
have been applied along the last years in quantum optics   
to study decoherence, by assuming a linear coupling between a
cavity harmonic mode and a thermal bath of oscillators at zero temperature,
as it has been done in \cite{Davidovitch} and \cite{Fonseca}. To investigate
decoherence of higher generation Schrodinger cat-states the cavity field
reduced matrix for these states could be calculated either by evaluating the
normal-ordering characteristic function, or by solving the evolution
equation for the field-resevoir state using the normal mode expansion,
generalizing the analysis of \cite{Davidovitch} and \cite{Fonseca}.

In this paper we adopt a
 general physicist's point of view, we do not intend to describe the 
specific features of a particular physical situation, instead   
we analyse a simplified linear version of the atom-field system and we try 
to extract the more detailed information we can from
this model. We take a linear simplified model in order to try to have a clearer
understanding of what we believe is one of the essential points, namely, the
need of non-perturbative analytical treatments to coupled systems,
which is the basic problem underlying the idea of dressed atom. Of
course, such an approach to a realistic non-linear system is an extremelly
hard task and here we make what we think is a good agreement between
physical reality and mathematical reliability, with the hope that in future
work our approach could be transposed to more realistic situations.

We consider a non
relativistic system composed of a harmonic oscillator coupled linearly to a
scalar field in ordinary Euclidean $3$-dimensional space. We start from an
analysis of the same system confined in a reflecting sphere of radius $R$,
and we assume that the free space solution to the radiating  oscillator
should be obtained taking a radius arbitrarily large in the $R$-dependent
quantities. The limit of an arbitrarily large radius in the mathematics of
the confined system is taken as a good description of the ordinary situation
of the radiating oscillator in free space. We will see that this is not
equivalent to the alternative continuous formulation in terms of
distributions, which is the case when we consider {\it a} {\it priori} the
system in unlimited space. The limiting procedure adopted here allows to
avoid the inherent ambiguities present in the continuous formulation. From a
physical point of view we give a non-perturbative treatment to the
oscillator radiation introducing some coordinates that allow to divide the
coupled system into two parts, the "dressed" oscillator and the field, what
makes unecessary to work directly with the concepts of "bare" oscillator,
field and interaction to study the radiation process. These are the main
reasons why we study a simplified linear system instead of a more realistic model,
to make evident some subtleties of the mathematics involved in the limiting
process of taking a cavity arbitrarily large, and also to exhibit an
exact solution valid for weak as well as for strong coupling. 
These aspects would be masked
in the perturbative approach used to study non-linear couplings. 

We start considering a harmonic oscillator $q_{0}(t)$ of frequency $%
\omega_{0}$ coupled linearly to a scalar field $\phi({\bf r},t)$, the whole
system being confined in a sphere of radius $R$ centered at the oscillator
position. The equations of motion are, 
\begin{equation}
\mbox{\"q}_{0}(t)+\omega_{0}^{2}q_{0}(t) = 2\pi \sqrt{gc}\int_{0}^{R}d^{3}%
{\bf r}\phi({\bf r},t)\delta ({\bf r})  \label{eq. mov1}
\end{equation}
\begin{equation}
\frac{1}{c^{2}}\frac{\partial^{2}\phi}{\partial t^{2}}-\nabla^{2} \phi({\bf r%
},t)=2\pi\sqrt{gc}q_{0}(t)\delta({\bf r})  \label{eq. mov2}
\end{equation}
which, using a basis of spherical Bessel functions defined in the domain $<|%
{\bf r}|<R$, may be written as a set of equations coupling the oscillator to
the harmonic field modes, 
\begin{equation}
\mbox{\"q}_{0}(t)+\omega_{0}^{2}q_{0}(t)=\eta\sum_{i=1}^{\infty}%
\omega_{i}q_{i}(t)  \label{eq. mov3.}
\end{equation}
\begin{equation}
\mbox{\"q}_{i}(t)+\omega_{i}^{2}q_{i}(t)=\eta\omega_{i}q_{0}(t).
\label{eq. mov4.}
\end{equation}
In the above equations, $g$ is a coupling constant, $\eta=\sqrt{%
2g\Delta\omega}$ and $\Delta\omega=\pi c/R$ is the interval between two
neighbouring field frequencies, $\omega_{i+1}-\omega_{i}=\Delta\omega=\pi c/R
$.

\section{The transformation to principal axis and the eigenfrequencies
spectrum}

{\bf 2.1 - Coupled harmonic Oscillators}

Let us consider for a moment the problem of a harmonic oscillator $q_{0}$
coupled to $N$ other oscillators. In the limit $N\rightarrow \infty$
we recover our original situation of the coupling oscillator-field  after
redefinition of divergent quantities, in a manner analogous as
renormalization is done in field theories. In terms of the cutoff $N$ the
coupled equations (\ref{eq. mov3.}) and (\ref{eq. mov4.}) are simply
rewritten taking the upper limit $N$ instead of $\infty$ for the summation
in the right hand side of Eq.(\ref{eq. mov3.}) and the system of $N+1$
coupled oscillators $q_{0}$ $\{q_{i}\}$ corresponds to the Hamiltonian, 
\begin{equation}
H=\frac{1}{2}\left[p_{0}^{2}+\omega_{0}^{2}q_{0}^{2}+
\sum_{k=1}^{N}p_{k}^{2}+\omega_{k}^{2}q_{k}^{2}-2\eta\omega_{k}q_{0}q_{k}%
\right].  \label{Hamiltoniana}
\end{equation}
The Hamiltonian (\ref{Hamiltoniana}) can be turned to principal axis by
means of a point tranformation, 
\begin{equation}
q_{\mu}=t_{\mu}^{r}Q_{r}\,,\,\,\,p_{\mu}=t_{\mu}^{r}P_{r},
\label{transformacao}
\end{equation}
performed by an orthonormal matrix $T=(t_{\mu}^{r})$, \, $\mu=(0,k)$, \, $%
k=1,2,...\, N$, $r=0,...N$. The subscript $0$ and $k$ refer respectively to
the oscillator and the harmonic modes  of the field and $r$ refers to the
normal modes. The transformed Hamiltonian in principal axis is 
\begin{equation}
H=\frac{1}{2}\sum_{r=0}^{N}(P_{r}^{2}+\Omega_{r}^{2}Q_{r}^{2}),
\label{HamiltonianaP}
\end{equation}
where the $\Omega_{r}$'s are the normal frequencies corresponding to the
possible collective oscillation modes of the coupled system.

Using the coordinate transformation $q_{\mu}=t_{\mu}^{r}Q_{r}$ in the
equations of motion and explicitly making use of the normalization condition 
$\sum_{\mu=0}^{N}(t_{\mu}^{r})^{2}=1$,  we get, 
\begin{equation}
t_{k}^{r}=\frac{\eta\omega_{k}}{\omega_{k}^{2}-\Omega_{r}^{2}}t_{0}^{r},
\label{tkr1}
\end{equation}

\begin{equation}
t_{0}^{r}= \left[1+\sum_{k=1}^{N}\frac{\eta^{2}\omega_{k}^{2}}{%
(\omega_{k}^{2}-\Omega_{r}^{2})^{2}}\right]^{-\frac{1}{2}}  \label{t0r}
\end{equation}

and 
\begin{equation}
\omega_{0}^{2}-\Omega_{r}^{2}=\eta^{2}\sum_{k=1}^{N}\frac{\omega_{k}^{2}}{%
\omega_{k}^{2}-\Omega_{r}^{2}}.  \label{tkr2}
\end{equation}
There are $N+1$ solutions $\Omega_{r}$ to Eq.(\ref{tkr2}), corresponding to
the $N+1$ normal collective oscillation modes. To have some insight into
these solutions, we take $\Omega_{r}=\Omega$ in Eq.(\ref{tkr2}) and
transform the right hand term. After some manipulations we obtain 
\begin{equation}
\omega_{0}^{2}-N\eta^{2}-\Omega^{2}=\eta^{2}\sum_{k=1}^{N}\frac{\Omega^{2}}{%
\omega_{k}^{2} -\Omega^{2}}  \label{Nelson1}
\end{equation}
It is easily seen that if $\omega_{0}^{2}>N\eta^{2}$ Eq.(\ref{Nelson1})
yelds only positive solutions for $\Omega^{2}$, what means that the system
oscillates harmonically in all its modes. Indeed, in this case the left hand
term of Eq.(\ref{Nelson1}) is positive for negative values of $\Omega^{2}$.
Conversely the right hand term is negative for those values of $\Omega^{2}$.
Thus there is no negative solution of that equation when $%
\omega_{0}^{2}>N\eta^{2}$. On the other hand it can be shown that if $%
\omega_{0}^{2}<N\eta^{2}$, Eq.(\ref{Nelson1}) has a single negative solution 
$\Omega_{-}^{2}$. In order to prove it let us define the function 
\begin{equation}
I(\Omega^{2})=(\omega_{0}^{2}-N\eta^{2})-\Omega^{2}-\eta^{2}\sum_{k=1}^{N} 
\frac{\Omega^{2}}{\omega_{k}^{2}-\Omega^{2}}  \label{Nelson2}
\end{equation}
Accordingly Eq.(\ref{Nelson1}) can be rewritten as $I(\Omega^{2})=0$. It can
be noticed that $I(\Omega^{2})\rightarrow \infty$ as $\Omega^{2}\rightarrow
-\infty$ and 
\begin{equation}
I(\Omega^{2}=0)=\omega_{0}^{2}-N\eta^{2}<0  \label{Nelson3}
\end{equation}
Furthermore $I(\Omega^{2})$ is a monotonically decreasing function in that
interval. Consequently $I(\Omega^{2})=0$ has a single negative solution when 
$\omega_{0}^{2}<N\eta^{2}$ as we have pointed out.  This means that there is
an oscillation mode whose amplitude varies exponentially and that does not
allows stationary configurations. We will not care about this last
situation. Thus we assume $\omega_{0}^{2}>N\eta^{2}$ and define the {\it %
renormalized} oscillator frequency $\bar{\omega}$ \cite{Thirring}, 
\begin{equation}
\bar{\omega}=\sqrt{\omega_{0}^{2}-N\eta^{2}}.  \label{omegabarra}
\end{equation}
In terms of the renormalized frequency Eq.(\ref{tkr2}) becomes, 
\begin{equation}
\bar{\omega}^{2}-\Omega_{r}^{2}=\eta^{2}\sum_{k=1}^{N}\frac{\Omega_{r}^{2}}{%
\omega_{k}^{2}-\Omega_{r}^{2}}.  \label{tkr3}
\end{equation}
From Eqs. (\ref{tkr1}), (\ref{t0r}) and (\ref{tkr3}), a straightforward
calculation shows the orthonormality relations for the transformation matrix 
$(t_{\mu}^{r})$. 

We get the transformation matrix elements for the oscillator-field system by
taking the limit $N\rightarrow \infty$ in the above equations. Recalling
the definition of $\eta$ from Eqs. (\ref{eq. mov3.}) and (\ref{eq. mov4.}),
we obtain after some algebraic manipulations, from Eqs. (\ref{tkr3}), (\ref
{tkr1}) and (\ref{t0r}), the matrix elements in the limit $N\rightarrow
\infty$, 
\begin{equation}
t_{0}^{r}=\frac{\Omega_{r}}{\sqrt{\frac{R}{2\pi gc}(\Omega_{r}^{2}-\bar{%
\omega}^{2})^{2}+\frac{1}{2}(3\Omega_{r}^{2}-\bar{\omega})^{2}+\frac{\pi gR}{%
2c}\Omega_{r}^{2}}}  \label{t0r1}
\end{equation}
and 
\begin{equation}
t_{k}^{r}=\frac{\eta\omega_{k}}{\omega_{k}^{2}-\Omega_{r}^{2}}t_{0}^{r}.
\label{t0r2}
\end{equation}

{\bf 2.2 - The eigenfrequencies spectrum}

Let us now return to the coupling oscillator-field by taking the limit $%
N\rightarrow \infty$ in the relations of the preceeding subsection. In this
limit it becomes clear the need for the frequency renormalization in Eq.(\ref
{omegabarra}). It is exactly the analogous of a mass renormalization in
field theory, the infinite $\omega_{0}$ is chosen in such a way as to make
the renormalized frequency $\bar{\omega}$ finite. Remembering Eq.(\ref{tkr3}) 
the solutions with respect
to the variable $\Omega$ of the equation  
\begin{equation}
\bar{\omega}^{2}-\Omega^{2}=\frac{2\pi gc}{R}\sum_{k=1}^{\infty}\frac{%
\Omega^{2}}{\omega_{k}^{2}-\Omega^{2}},  \label{tkr6}
\end{equation}
give the collective modes frequencies. We remember $\omega_{k}= k\frac{\pi c%
}{R}$, $k=1,2,...$, and take a positive $x$ such that $\Omega= x\frac{\pi c}{%
R}$. Then using the identity, 
\begin{equation}
\sum_{k=1}^{\infty}\frac{x^{2}}{k^{2}-\Omega^{2}}=\frac{1}{2}(1-\pi x\cot\pi
x),  \label{cotg1}
\end{equation}
Eq.(\ref{tkr6}) may be rewritten in the form, 
\begin{equation}
cotg\pi x=\frac{c}{Rg}x+\frac{1}{\pi x}(1-\frac{R\bar{\omega}^{2}}{\pi gc}).
\label{cotg3}
\end{equation}
The secant curve corresponding to the right hand side of the above equation
cuts only once each branch of the cotangent in the left hand side. Thus we
may label the solutions $x_{r}$ as $x_{r}=r+\epsilon_{r}$, $0<\epsilon_{r}<1$%
, $r=0,1,2...$, and the collective eigenfrequencies are, 
\begin{equation}
\Omega_{r}=(r+\epsilon_{r})\frac{\pi c}{R},  \label{Omega}
\end{equation}
the $\epsilon$'s satisfying the equation, 
\begin{equation}
cot(\pi \epsilon_{r})=\frac{\Omega_{r}^{2}-\bar{\omega}^{2}}{\Omega_{r}\pi g}%
+\frac{c}{\Omega_{r}R}.  
\label{epsilon}
\end{equation}

The field $\phi({\bf r},t)$ can be expressed in terms of the normal modes.
We start from its expansion in terms of spherical Bessel functions, 
\begin{equation}
\phi({\bf r},t)=c\sum_{k=1}^{\infty}q_{k}(t)\phi_{k}({\bf r}),
\label{Fourier1}
\end{equation}
where 
\begin{equation}
\phi_{k}({\bf r})=\frac{sin\frac{\omega_{k}}{c}|{\bf r}|}{{\bf r}\sqrt{2\pi R%
}}.  \label{Fourier2}
\end{equation}
Using the principal axis transformation matrix together with the equations
of motion we obtain an expansion for the field in terms of an orthonormal
basis associated to the collective normal modes, 
\begin{equation}
\phi({\bf r},t)=c\sum_{s=0}^{\infty}Q_{s}(t)\Phi_{s}({\bf r}),
\label{Fourier3}
\end{equation}
where the normal collective Fourier modes 
\begin{equation}
\Phi_{s}({\bf r})=\sum_{k}t_{k}^{s}\frac{sin\frac{\omega_{k}}{c}|{\bf r}|}{%
{\bf r}\sqrt{2\pi R}}  
\label{Fourier4}
\end{equation}
satisfy the equation 
\begin{equation}
(-\frac{\Omega_{s}^{2}}{c^{2}}-\Delta)\phi_{s}({\bf r})=2\pi \sqrt{\frac{g}{c%
}}\delta({\bf r})t_{0}^{s},  
\label{Fourier5}
\end{equation}
which has a solution of the form 
\begin{equation}
\phi({\bf r},t)=-\sqrt{\frac{g}{c}}\frac{t_{0}^{s}}{2|{\bf r}|sin\delta_{s}}%
sin(\frac{\Omega_{s}}{c}|{\bf r}|-\delta_{s}).  
\label{Fourier6}
\end{equation}
To determine the phase $\delta_{s}$ we expand the right hand term of Eq.(\ref
{Fourier6}) and compare with the formal expansion (\ref{Fourier4}). This
imply the condition 
\begin{equation}
sin(\frac{\Omega_{s}}{c}R-\delta_{s})=0.  
\label{Fourier5'}
\end{equation}
Remembering from Eq.(\ref{Omega}) that there is $0<\epsilon_{s}<1$ such that 
$\Omega_{s}=(s+\epsilon_{s})\frac{\pi}{R}$, it is easy to show from the
condition in Eq.(\ref{Fourier5}) that the phase $0<\delta_{s}<\pi$ has the
form 
\begin{equation}
\delta_{s}=\epsilon_{s}\pi.  \label{fase}
\end{equation}
Comparing Eqs.(\ref{Fourier2}) and (\ref{Fourier4}) and using the explicit
form (\ref{t0r1}) of the matrix element $t_{0}^{s}$ we obtain the expansion
for the field in terms of the normal collective modes, 
\begin{equation}
\phi({\bf r},t)=-\frac{\sqrt{gc}}{2}\sum_{s}\frac{Q_{s}sin(\frac{\Omega_{s}}{%
c}|{\bf r}|-\delta_{s})}{|{\bf r}|\sqrt{sin^{2}\delta_{s}+(\frac{\eta R}{2c}%
)^{2}(1-\frac{sin \delta_{s}cos \delta_{s}}{\Omega_{s} R/c}})}  \label{campo}
\end{equation}

\section{The limit $R\rightarrow \infty$ - mathematical aspects}

{\bf 3.1 - Discussion of the mathematical problem}

Unless
explicitly stated, in the remaining of this paper the symbol $R\rightarrow
\infty$ is to be understood as the situation of a cavity of fixed,
arbitrarily large radius.
In order to compare the behaviour of the system in a very large cavity to
that it would be in free space, let us firstly consider the system embedded
in an {\it a} {\it priori} infinite Euclidean space; in this case to compute
the quantities describing the system means essentially to replace by
integrals the discrete sums appearing in the confined problem, taking
direcltly $R=\infty$. An alternative procedure is to compute the quantities
describing the system confined in a sphere of radius $R$ and take the limit $%
R\rightarrow \infty$ afterwards. This last approach to describe the system
in free space should keep in some way the "memory" of the confined system.
To be physically equivalent one should expect that the two approachs give
the same results. We will see that at least from a mathematical point of
view this is not exactly the case. We remark that a solution
to the problem of a system composed of an oscillator coupled 
to a field in free space, is already known since a long time
ago \cite{Ullersma} in the context of Bownian motion. This solution is quite
different from ours, in the sense that it not concerns the system confined to a box
and also that it is limited to the dipole term from the multipolar expansion to
the field.   

In the continuous formalism of free space the field normal modes Fourier
components (analogous to the components $\phi_{s}$ in Eq.(\ref{Fourier4}))
are, 
\begin{equation}
\phi_{\Omega}=h(\Omega)\int_{0}^{\infty}\,d\omega \frac{\omega}{%
\omega^{2}-\Omega^{2}}\frac{sin\frac{\omega}{c}|{\bf r}|}{|{\bf r}|},
\label{Fourier continua 1}
\end{equation}
where 
\begin{equation}
h(\Omega)=\frac{2g\Omega}{\sqrt{(\Omega^{2}-\bar{\omega}^{2})^{2}+\pi
g^{2}\Omega^{2}}}  \label{Fourier continua 2}
\end{equation}
and where the we have taken the appropriate continuous form of Eqs.(\ref
{t0r1}) and (\ref{t0r2}). Splitting $\omega/(\omega^{2}-\Omega^{2})$ into
partial fractions we get 
\begin{equation}
\phi_{\Omega}=h(\Omega)\int_{-\infty}^{+\infty}\,d\omega \frac{1}{%
\omega-\Omega}\frac{sin\frac{\omega}{c}|{\bf r}|}{|{\bf r}|}.
\label{Fourier continua 3}
\end{equation}

The pole at $\omega=\Omega$ prevents the existence of the integral in
Eq.(\ref{Fourier continua 3}). The usual way to circumvect this difficulty
is to replace the integral by one of the quantities, 
\begin{equation}
Lim_{\epsilon \rightarrow 0}\int_{-\infty}^{+\infty}\,d\omega \frac{1}{%
\omega-(\Omega \pm i\epsilon)}\frac{sin\frac{\omega}{c}|{\bf r}|}{|{\bf r}|}%
\equiv \int_{-\infty}^{+\infty}\,d\omega \delta_{\pm}(\omega-\Omega)\frac{sin%
\frac{\omega}{c}|{\bf r}|}{|{\bf r}|} ,  \label{distribuicao 1}
\end{equation}
where 
\begin{equation}
\delta_{\pm}(\omega-\Omega)=\frac{1}{\pi}P(\frac{1}{\omega-\Omega})\pm
i\delta (\omega-\Omega),  \label{distribuicao 2}
\end{equation}
with $P$ standing for principal value. In our case this redefinition of the
normal modes Fourier components may be justified by the fact that both
integrals in Eq.(\ref{distribuicao 1}) are solutions of the equations of
motion (\ref{eq. mov1}) and (\ref{eq. mov2}) for ${\bf r}\neq 0$, and so the
solution should be a linear combination of them. The situation is different
if we adopt the point of view of taking the limit $R\rightarrow \infty$ in
the solution of the confined problem. In this case the Fourier component $%
\phi_{\Omega}$ is obtained by taking the limit $R\rightarrow \infty$ in the
expression for the field, Eq(\ref{Fourier6}), what allows to obtain an
uniquely defined expression to the normal modes Fourier components, to each $%
\phi_{\Omega}$ corresponding a phase $\delta_{\Omega}$ (the limit $%
R\rightarrow \infty$ of $\delta_{s}$ in Eq.(\ref{epsilon}) given by 
\begin{equation}
cot\delta_{\Omega}=\frac{1}{\pi g}\frac{\Omega^{2}-\bar{\omega}^{2}}{\Omega}.
\label{fase continua}
\end{equation}
Also, comparing Eqs.(\ref{distribuicao 1}), (\ref{distribuicao 2}) and (\ref
{Fourier4}) we see that the adoption of the continuous formalism is
equivalent to assume that in the limit $R\rightarrow \infty$ the elements $%
t_{i}^{s}$ of the transformation matrix should be replaced by $%
\delta_{+}(\omega-\Omega)$ or by $\delta_{-}(\omega-\Omega)$. This procedure
is, from a mathematical point of view, perfectly justified but at the price
of loosing uniqueness in the definition of the field components.

If we take the solution of the confined problem  and we compute the matrix
elements $t_{i}^{s}$ for $R$ arbitrarily large, we will see in subsection 
{\bf 3.2} that these elements do not tend to distributions in this limit. As 
$R$ becomes larger and larger the set of non-vanishing elements $t_{i}^{s}$
concentrate for each $i$ in a small neighbourhood of $\omega_{i}$. In the
limit $R\rightarrow \infty$ the whole set of the matrix elements $t_{i}^{s}$
contains an arbitrarily large number of elements quadratically summables 
\cite{Beck}. For the matrix elements $t_{0}^{s}$ we obtain a quadratically
integrable expression.

In the continuous formulation the unit matrix, corresponding to the absence
of coupling, has elements $E_{\omega}^{\Omega}=\delta(\omega-\Omega)$, while
if we start from the confined situation, it can be verified that in the
limit $g\rightarrow 0$, $R\rightarrow \infty$, the matrix $T=(t_{\mu}^{s})$
tends to the usual unit matrix of elements $E_{\omega
,\Omega}=\delta_{\omega,\Omega}$.

The basic quantity describing the system, the transformation matrix $%
T=(t_{\mu}^{s})$ has, as we will see, different properties in free space, if
we use the continuous formalism or if we adopt the procedure of taking the
limit $R\rightarrow \infty$ from the matrix elements in the confined problem
. In the first case we must define the matrix elements $t_{\omega}^{\Omega}$
linking free field modes to normal modes, as distributions. On the other
side adopting the second procedure we will find that the
limiting matrix elements $Lim_{R\rightarrow \infty}\, t_{i}^{s}$ are not
distributions, but well defined finite quantities. The two procedures are
not equivalent, the limit $R\rightarrow \infty$ does not commute with other
operations. In this note we take as physically meaningfull the second
procedure, we solve first the problem in the confined case (finite $R$) and
take afterwards the limit of infinite (in the sense of arbitrarily large)
radius of the cavity. In the next subsection we perform a detailed analysis
of the limit $R\rightarrow \infty$ of the transformation matrix $%
(t_{\mu}^{r})$.

{\bf 3.2 - The transformation matrix in the limit $R\rightarrow \infty$}

From Eqs. (\ref{t0r1}) and (\ref{t0r2}) we obtain for $R$ arbitrarily large, 
\begin{equation}
t_{0}^{r}\rightarrow Lim_{\Delta \Omega \rightarrow 0}\, t_{\bar{\omega}%
}^{\Omega}\sqrt{\Delta \Omega}=Lim_{\Delta \Omega \rightarrow 0}\frac{\sqrt{%
2g}\Omega \sqrt{\Delta \Omega}}{\sqrt{(\Omega^{2}-\bar{\omega}%
^{2})^{2}+\pi^{2}g^{2}\Omega^{2}}}.  \label{t0r(R)}
\end{equation}
and 
\begin{equation}
t_{k}^{r}=\frac{2g\omega_{k}\Delta \omega}{(\omega_{k}+\Omega_{r})(%
\omega_{k}-\Omega_{r})}\frac{\Omega_{r}}{\sqrt{(\Omega_{r}^{2}-\bar{\omega}%
^{2})^{2}+\pi^{2}g^{2}\Omega_{r}^{2}}},  \label{tir(R)}
\end{equation}
where we have used the fact that in this limit $\Delta \omega=\Delta \Omega=%
\frac{\pi c}{R}$. The matrix elements $t_{\bar{\omega}}^{\Omega}$ are
quadratically integrable to one, $\int(t_{\bar{\omega}}^{\Omega})^{2}\,d%
\Omega=1$, as may be seen using Cauchy theorem.

For $R$ arbitrarily large ($\Delta \omega=\frac{\pi c}{R}\rightarrow 0$),
the only nonvanishing matrix elements $t_{i}^{r}$ are those for which $%
\omega_{i}-\Omega_{r}\approx \Delta \omega$. To get explicit formulas for
these matrix elements in the limit $R\rightarrow \infty$ let us consider $R$
large enough such that we may take $\Delta \omega \approx \Delta \Omega$ and
consider the points of the spectrum of eigenfrequencies $\Omega$ inside and
outside a neighbourhood $\eta$ (defined in Eqs.(\ref{eq. mov3.}) and (\ref
{eq. mov4.}) of $\omega_{i}$. We note that $R>\frac{2\pi c}{g}$ implies $%
\frac{\eta}{2}>\Delta \omega$, then we may consider $R$ such that the right
(left) neighbourhood $\frac{\eta}{2}$ of $\omega_{i}$ contains an integer
number, $\kappa$, of frequencies $\Omega_{r}$, 
\begin{equation}
\kappa \Delta \omega =\frac{\eta}{2}=\sqrt{\frac{g\Delta \omega}{2}}.
\label{kappa}
\end{equation}
If $R$ is arbitrarily large we see from (\ref{kappa}) that $\frac{\eta}{2}$
is arbitrarily small, but $\kappa$ grows at the same rate, what means
firstly that the difference $\omega_{i}-\Omega_{r}$ for the $\Omega_{r}$'s
outside the neighbourhood $\eta$ of $\omega_{i}$ is abitrarily larger than $%
\Delta \omega$, implying that the corresponding matrix elements $t_{i}^{r}$
tend to zero (see Eq.(\ref{tir(R)})). Secondly all frequencies $\Omega_{r}$
inside the neighbourhood $\eta$ of $\omega_{i}$ are arbitrarily close to $%
\omega_{i}$, being in arbitrarily large number. Only the matrix  elements $%
t_{i}^{r}$ corresponding to these frequencies $\Omega_{r}$ inside the
neighbourhood $\eta$ of $\omega_{i}$ are different from zero. For these we
make the change of labels, 
\begin{equation}
r=i-n \, (\omega_{i}-\frac{\eta}{2}<\Omega_{r}<\omega_{i})\, ;\, r=i+n \,
(\omega_{i}>\Omega_{r}>\omega_{i}+\frac{\eta}{2}),  \label{labels}
\end{equation}
$i=1,2,...$. We get, from Eq.(\ref{tir(R)}) 
\begin{equation}
t_{i}^{i}=\frac{g\omega_{i}}{\sqrt{(\Omega_{r}^{2}-\bar{\omega}%
^{2})^{2}+\pi^{2}g^{2}\omega_{i}^{2}}}\frac{1}{\epsilon_{i}}  \label{tir1(R)}
\end{equation}
and 
\begin{equation}
t_{i}^{i\pm n}=\frac{\mp g\omega_{i}}{\sqrt{(\Omega_{r}^{2}-\bar{\omega}%
^{2})^{2}+\pi^{2}g^{2}\omega_{i}^{2}}}\frac{1}{n\pm \epsilon_{i}},
\label{tir2(R)}
\end{equation}
where $\epsilon_{i}$ satisfies Eq.(\ref{epsilon}) in this case, 
\begin{equation}
cot(\pi \epsilon_{i})=\frac{\omega_{i}^{2}-\bar{\omega}^{2}}{\omega_{i}\pi g}%
.  \label{epsilon1}
\end{equation}
Using the formula 
\begin{equation}
\pi^{2}cosec^{2}(\pi \epsilon_{i})=\frac{1}{\epsilon_{i}}+\sum_{n=1}^{\infty}%
\left[\frac{1}{(n+\epsilon_{i})^{2}}+\frac{1}{(n-\epsilon_{i})^{2}}\right],
\label{epsilon2}
\end{equation}
it is easy to show the normalization condition for the matrix elements (\ref
{tir1(R)}) and (\ref{tir2(R)}), 
\begin{equation}
(t_{i}^{i})^{2}+\sum_{n=1}^{\infty}(t_{i}^{i-n})^{2}+(t_{i}^{i+n})^{2}=1
\label{tir3(R)}
\end{equation}
and also the orthogonality relation, 
\begin{equation}
\sum_{r}t_{i}^{r}t_{k}^{r}=0 \, (i\neq k)  \label{tir4(R)}
\end{equation}
in the limit $R\rightarrow \infty$.

{\bf 3.3 - The transformation matrix in the limit $g=0$}

From Eq. (\ref{t0r1}) we get for arbitrary $R$, 
\begin{equation}
Lim_{g\rightarrow 0} \, t_{0}^{r}=\cases {1,&if $\Omega_{r}=\bar{\omega}$;
\cr 0,&otherwise.\cr}.  \label{t0r(g=0)}
\end{equation}
From Eqs.(\ref{tir1(R)}) and (\ref{tir2(R)}) we see that the matrix elements 
$t_{i}^{r}$ for $i\neq r$ all vanish for $g=0$. Also, using Eqs.(\ref{Omega}%
) we obtain for small $g$, 
\begin{equation}
t_{i}^{i}\approx \frac{2g\Omega_{i}\omega_{i}}{(\Omega_{i}^{2}-\bar{\omega}%
^{2})(\omega_{i}+\Omega_{i})}\frac{1}{\epsilon_{i}},  \label{tir(g pequeno)}
\end{equation}
or, expanding $\epsilon_{i}$ for small $g$ from Eq.(\ref{epsilon1}) 
\begin{equation}
t_{i}^{i}(g=0)=1  \label{tir(g=0)}
\end{equation}

We see from the above expressions that in the limit $R\rightarrow \infty$
the matrix $(t_{\mu}^{r})$ remains an orthonormal matrix in the usual sense
as for finite $R$. With the choice of the procedure of taking the limit $%
R\rightarrow \infty$ from the confined solution, the matrix elements do not
tend to distributions in the free space limit  as it would be the case using
the continuous formalism. All non- vanishing matrix elements $t_{i}^{r}$ are
concentrated inside a neighbourhood $\eta$ of $\omega_{i}$, their set is a
quadratically summable  enumerable set. The elements $(t_{0}^{r})$ tend to a
quadratically integrable expression.

\section{The Radiation Process}

We start this section defining some coordinates $q^{\prime}_{0}$, $%
q^{\prime}_{i}$ associated  to the "dressed" mechanical oscillator and to
the field. These coordinates will reveal themselves to be suitable to give
an appealling non-perturbative description of the oscillator-field system.
The general conditions that such coordinates must satisfy, taking into
account that the system is rigorously described by the collective normal
coordinates modes $Q_{r}$, are the following:

- In reason of the linear character of our problem the coordinates $%
q^{\prime}_{0}$, $q^{\prime}_{i}$ should be linear functions of the
collective coordinates $Q_{r}$

- They should allow to construct ortogonal configurations corresponding to
the separation of the system into two parts, the dressed oscillator and the
field.

- The set of these configurations should contain the ground state, $%
\Gamma_{0}$.

The last of the above conditions restricts the transformation between the
coordinates $q^{\prime}_{\mu}$, $\mu=0,i=1,2,...$ and the collective ones $%
Q_{r}$ to those leaving invariant the quadratic form, 
\begin{equation}
\sum_{r}\Omega_{r}Q_{r}^{2}=\bar{\omega}(q^{\prime}_{0})^{2}+\sum_{i}%
\omega_{i}(q^{\prime}_{i})^{2}\,  \label{quadratica}
\end{equation}
Our configurations will behave in a first approximation as independent
states, but they will evolve as the time goes on, as if transitions among
them were being in progress, while the basic configuration $\Gamma_{0}$
represents a rigorous eigenstate of the system and does not change with
time. The new coordinates $q^{\prime}_{\mu}$ describe dressed configurations
of the oscillator and field quanta.

{\bf 4.1 - The dressed coordinates $q^{\prime}_{\mu}$}

The eigenstates of our system are represented by the normalized
eigenfunctions, 
\begin{equation}
\phi_{n_{0}n_{1}n_{2}...}(Q,t)=\prod_{s}\left[N_{n_{s}}H_{n_{s}}(\sqrt{\frac{%
\Omega_{s}}{\hbar}}Q_{s})\right]\Gamma_{0}e^{-i\sum_{s}n_{s}\Omega_{s}t},
\label{autofuncoes}
\end{equation}
where $H_{n_{s}}$ is the $n_{s}$-th Hermite polynomial, $N_{n_{s}}$ is a
normalization coefficient, 
\begin{equation}
N_{n_{s}}=(2^{-n_{s}}n_{s}!)^{-\frac{1}{2}}  \label{normalizacao}
\end{equation}
and $\Gamma_{0}$ is a normalized representation of the ground state, 
\begin{equation}
\Gamma_{0}=exp\left[-\sum_{s}\frac{\Omega_{s}Q_{s}^{2}}{2\hbar}-\frac{1}{4}ln%
\frac{\Omega_{s}}{\pi \hbar}\right].  \label{vacuo1}
\end{equation}

To describe the radiation process, having as initial condition that only the
mechanical oscillator, $q_{0}$ be excited, the usual procedure is to
consider the interaction term in the Hamiltonian written in terms of $q_{0}$%
, $q_{i}$ as a perturbation, which induces transitions among the eigenstates
of the free Hamiltonian. In this way it is possible to treat approximatelly
the problem having as initial condition that only the bare oscillator be
excited. But as is well known this initial condition is physically not
consistent due to the divergence of the bare oscillator frequency if there
is interaction with the field. The traditional way to circumvect this
difficulty is by the renormalization procedure, introducing perturbativelly
order by order corrections to the oscillator frequency. Here we adopt an
alternative procedure, we do not make explicit use of the concepts of
interacting bare oscillator and field, described by the coordinates $q_{0}$
and $\{q_{i}\}$, we introduce "dressed" coordinates $q^{\prime}_{0}$ and $%
\{q^{\prime}_{i}\}$ for, respectivelly the "dressed" oscillator and the
field, defined by, 
\begin{equation}
\sqrt{\frac{\bar{\omega}_{\mu}}{\hbar}}q^{\prime}_{\mu}=\sum_{r}t_{\mu}^{r}%
\sqrt{\frac{\Omega_{r}}{\hbar}}Q_{r},  \label{qvestidas1}
\end{equation}
valid for arbitrary $R$, which satisfy the condition to leave invariant the
quadratic form (\ref{quadratica}) and where $\bar{\omega}_{\mu}=\bar{\omega}%
,\, \{\omega_{i}\}$. In terms of the bare coordinates the dressed
coordinates are expressed as, 
\begin{equation}
q^{\prime}_{\mu}=\sum_{\nu}\alpha_{\mu \nu}q_{\nu},  \label{qvestidas3}
\end{equation}
where 
\begin{equation}
\alpha_{\mu \nu}=\frac{1}{\sqrt{\bar{\omega}_{\mu}}}\sum_{r}t_{\mu}^{r}t_{%
\nu}^{r}\sqrt{\Omega_{r}}.  \label{qvestidas4}
\end{equation}

As $R$ becomes larger and larger we get for the various coefficients $\alpha$
in Eq.(\ref{qvestidas4}):

a) from Eq.(\ref{t0r(R)}), 
\begin{equation}
Lim_{R\rightarrow \infty} \, \alpha_{00}=\frac{1}{\sqrt{\bar{\omega}}}%
\int_{0}^{\infty} \frac{2g\Omega^{2}\sqrt{\Omega} d\Omega}{(\Omega^{2}-\bar{%
\omega}^{2})^{2}+\pi^{2}g^{2}\Omega^{2}}\equiv A_{00}(\bar{\omega},g).
\label{alfa00}
\end{equation}
b) To evaluate $\alpha_{0i}$ and $\alpha_{0i}$ in the limit $R\rightarrow
\infty$, we remember from the discussion in subsection {\bf 3.2} that in the
the limit $R\rightarrow \infty$, for each $i$  the only non-vanishing matrix
elements $t_{i}^{r}$ are those for which the corresponding eigenfrequencies $%
\Omega_{r}$ are arbitrarily near the field frequency $\omega_{i}$. We obtain
from Eqs. (\ref{t0r(R)}), (\ref{tir1(R)}) and (\ref{tir2(R)}), 
\begin{equation}
Lim_{R\rightarrow \infty} \, \alpha_{i0}=Lim_{\Delta \omega \rightarrow 0}%
\frac{1}{\sqrt{\omega_{i}}}\frac{(2g^{2}\omega_{i}^{5}\Delta \omega)^{\frac{1%
}{2}}} {(\omega_{i}^{2}-\bar{\omega}^{2})^{2}+\pi^{2}g^{2}\omega_{i}^{2}}%
(\sum_{n=1}^{\infty} \frac{2\epsilon_{i}}{n^{2}-\epsilon_{i}^{2}}-\frac{1}{%
\epsilon_{i}})  \label{alfai0}
\end{equation}
and 
\begin{equation}
Lim_{R\rightarrow \infty} \, \alpha_{0i}=Lim_{\Delta \omega \rightarrow 0}%
\frac{1}{\sqrt{\bar{\omega}}}\frac{(2g^{2}\omega_{i}^{5}\Delta \omega)^{%
\frac{1}{2}}} {(\omega_{i}^{2}-\bar{\omega}^{2})^{2}+\pi^{2}g^{2}%
\omega_{i}^{2}}(\sum_{n=1}^{\infty} \frac{2\epsilon_{i}}{n^{2}-%
\epsilon_{i}^{2}}-\frac{1}{\epsilon_{i}})  \label{alfa0i}
\end{equation}
c) Since in the limit $R\rightarrow \infty$ the only non-zero matrix
elements $t_{i}^{r}$ corresponds to $\Omega_{r}=\omega_{i}$, the product $%
t_{i}^{r}t_{k}^{r}$ vanishes for $\omega_{i}\neq \omega_{k}$. Then we obtain
from Eqs.(\ref{qvestidas4}) and (\ref{tir3(R)}) 
\begin{equation}
Lim_{R\rightarrow \infty} \, \alpha_{ik}=\delta_{ik}.  \label{alfaik}
\end{equation}
Thus, from Eqs.(\ref{qvestidas3}), (\ref{alfaik}), (\ref{alfai0}), (\ref
{alfa0i}) and (\ref{alfa00}) we can express the dressed coordinates $%
q^{\prime}_{\mu}$ in terms of the bare ones, $q_{\mu}$ in the limit $%
R\rightarrow \infty$, 
\begin{equation}
q^{\prime}_{0}=A_{00}(\bar{\omega},g)q_{0},  \label{q0'q0}
\end{equation}
\begin{equation}
q^{\prime}_{i}=q_{i}.  \label{qi'qi}
\end{equation}

It is interesting to compare Eqs.(\ref{qvestidas3}) with Eqs.(\ref{q0'q0}), (%
\ref{qi'qi}). In the case of Eqs.(\ref{qvestidas3}) for finite $R$, the
coordinates $q^{\prime}_{0}$ and $\{q^{\prime}_{i}\}$ are all dressed, in
the sense that they are all collective, both the field modes and the
mechanical oscillator can not be separeted in this language. In the limit $%
R\rightarrow \infty$, Eqs.(\ref{q0'q0}) and (\ref{qi'qi}) tells us that the
coordinate $q^{\prime}_{0}$ describes the mechanical oscillator modified by
the presence of the field in a indissoluble way, the mechanical oscillator
is always dressed by the field. On the other side, the dressed harmonic
modes of the field, described by the coordinates $q^{\prime}_{i}$ are
identical to the bare field modes, in other words, the field keeps in the
limit $R\rightarrow \infty$ its proper identity, while the mechanical
oscillator is always accompanied by a cloud of field quanta. Therefore we
identify the coordinate $q^{\prime}_{0}$ as the coordinate describing the
mechanical oscillator dressed by its proper field, being the whole system
divided into dressed oscillator and field, without appeal to the concept of
interaction between them, the interaction being absorbed in the dressing
cloud of the oscillator. In the next subsections we use the dressed
coordinates to describe the radiation process.

{\bf 4.2 - Dressed configurations and the radiation process}

Let us define for a fixed instant the complete orthonormal set of functions, 
\begin{equation}
\psi_{\kappa_{0} \kappa_{1}...}(q^{\prime})=\prod_{\mu}\left[%
N_{\kappa_{\mu}}H_{\kappa_{\mu}} (\sqrt{\frac{\bar{\omega}_{\mu}}{\hbar}}%
q^{\prime}_{\mu})\right]\Gamma_{0},  \label{ortovestidas1}
\end{equation}
where $q^{\prime}_{\mu}=q^{\prime}_{0},\, q^{\prime}_{i}$, $\bar{\omega}%
_{\mu}=\bar{\omega},\, \omega_{i}$ and $N_{\kappa_{\mu}}$ and $\Gamma_{0}$
are as in Eq.(\ref{autofuncoes}). Using Eq.(\ref{qvestidas1}) the functions (%
\ref{ortovestidas1}) can be expressed in terms of the normal coordinates $%
Q_{r}$. But since (\ref{autofuncoes}) is a complete set of orthonormal
functions, the functions (\ref{ortovestidas1}) may be written as linear
combinations of the eigenfunctions of the coupled system (we take $t=0$ for
the moment), 
\begin{equation}
\psi_{\kappa_{0}
\kappa_{1}...}(q^{\prime})=\sum_{n_{0}n_{1}...}T_{\kappa_{0} \kappa_{1}...}
^{n_{0}n_{1}...}(0)\phi_{n_{0}n_{1}n_{2}...}(Q,0),  \label{ortovestidas2}
\end{equation}
where the coefficients are given by, 
\begin{equation}
T_{\kappa_{0} \kappa_{1}...}^{n_{0}n_{1}...}(0)=\int dQ\, \psi_{\kappa_{0}
\kappa_{1}...}\phi_{n_{0}n_{1}n_{2}...},  \label{ortovestidas3}
\end{equation}
the integral extending over the whole $Q$-space.

We consider the particular configuration $\psi$ in which only one dressed
oscillator $q^{\prime}_{\mu}$ is in its $N$-th excited state, 
\begin{equation}
\psi_{0...N(\mu)0...}(q^{\prime})=N_{N} H_{N}(\sqrt{\frac{\bar{\omega}_{\mu}%
}{\hbar}}q^{\prime}_{\mu})\Gamma_{0}.  \label{ortovestidas4}
\end{equation}
The coefficients (\ref{ortovestidas3}) can be calculated in this case using
Eqs.(\ref{ortovestidas3}), (\ref{ortovestidas1}) and (\ref{qvestidas1}) with
the help of the theorem \cite{Ederlyi}, 
\begin{equation}
\frac{1}{m!}\left[\sum_{r}(t_{\mu}^{r})^{2}\right]^{\frac{m}{2}}H_{N}(\frac{%
\sum_{r}t_{\mu}^{r}\sqrt{\frac{\Omega_{r}}{\hbar}}Q_{r}}{\sqrt{%
\sum_{r}(t_{\mu}^{r})^{2}}}) =\sum_{m_{0}+m_{1}+...=N}\frac{%
(t_{\mu}^{0})^{m_{0}}(t_{\mu}^{1})^{m_{1}}...}{m_{0}!m_{1}!...}H_{m_{0}}(%
\sqrt{\frac{\Omega_{0}}{\hbar}}Q_{0})H_{m_{1}}(\sqrt{\frac{\Omega_{1}}{\hbar}%
}Q_{1})...  \label{teorema Ederlyi}
\end{equation}
We get, 
\begin{equation}
T_{0...N(\mu)0...}^{n_{0}n_{1}...}=(\frac{m!}{n_{0}!n_{1}!...})^{\frac{1}{2}%
}(t_{\mu}^{0})^{n_{0}}(t_{\mu}^{1})^{n_{1}}...,  \label{coeffN}
\end{equation}
where the subscripts $\mu=0,\, i$ refer respectivelly to the dressed
mechanical oscillator and the harmonic modes of the field and the quantum
numbers are submited to the  constraint $n_{0}+n_{1}+...=N$.

In the following we study the behaviour of the system with the initial
condition that only the dressed mechanical oscillator $q^{\prime}_{0}$ be in
the $N$-th excited state. We will study in detail the particular cases $N=1$
and $N=2$, which will be enough to have a clear understanding of our
approach.

- $N=1$: Let us call $\Gamma_{1}^{\mu}$ the configuration in which only the
dressed oscillator $q^{\prime}_{\mu}$ is in the first excited level. The
initial configuration in which the dressed mechanical oscillator is in the
first excited level is $\Gamma_{1}^{0}$. We have from Eq.(\ref{ortovestidas4}%
), (\ref{ortovestidas2}) (\ref{coeffN}) and (\ref{qvestidas1}) the following
expression for the time evolution of the first-level excited dressed
oscillator $q^{\prime}_{\mu}$, 
\begin{equation}
\Gamma_{1}^{\mu}=\sum_{\nu}f^{\mu \nu}(t)\Gamma_{1}^{\nu}(0),
\label{ortovestidas5}
\end{equation}
where the coefficients $f^{\mu \nu}(t)$ are given by 
\begin{equation}
f^{\mu \nu}(t)=\sum_{s}t_{\mu}^{s}t_{\nu}^{s}e^{-i\Omega_{s}t},
\label{fmunu}
\end{equation}
That is, the initially excited dressed oscillator naturally distributes its
energy among itself and all others dressed oscillators, as time goes on. If
the mechanical dressed oscillator is in its first excited state at $t=0$,
its decay rate may evaluated from its time evolution equation, 
\begin{equation}
\Gamma_{1}^{0}=\sum_{\nu}f^{0 \nu}(t)\Gamma_{1}^{\nu}(0).
\label{ortovestidas6}
\end{equation}
In Eq.(\ref{ortovestidas6}) the coefficients $f^{0 \nu}(t)$ have a simple
interpretation: remembering Eqs.(\ref{q0'q0}) and (\ref{qi'qi}), $f^{00}(t)$
and $f^{0i}(t)$  are respectivelly the probability amplitudes that at time $t
$ the dressed mechanical oscillator still be excited or have radiated a
field quantum of frequency $\omega_{i}$. We see that this formalism allows a
quite natural description of the radiation process as a simple exact time
evolution of the system. Let us for instance evaluate the oscillator decay
probability in this language. From Eqs.(\ref{t0r(R)}) and (\ref{fmunu}) we
get 
\begin{equation}
f^{00}(t)=\int_{0}^{\infty}\frac{2g\Omega^{2}e^{-i\Omega t}\, d\Omega} {%
(\Omega^{2}-\omega^{2})^{2}+\pi^{2}g^{2}\Omega^{2}}.  \label{f00}
\end{equation}
The above integral can be evaluated by Cauchy theorem. For large $t$ ($t>>%
\frac{1}{\bar{\omega}}$), but arbitrary coupling $g$, we obtain for the
oscillator decay probability, the result, 
\begin{equation}
|f^{00}(t)|^{2}=e^{-\pi gt}(1+\frac{\pi^{2}g^{2}}{4\bar{\omega}^{2}}%
)+e^{-\pi gt} \frac{8\pi g}{\pi \bar{\omega}^{4}t^{3}}(sin\tilde{\bar{\omega}%
}t+\frac{\pi g}{2<\bar{\omega}>} cos\tilde{\bar{\omega}}t)+\frac{%
16\pi^{2}g^{2}}{\pi^{2}\bar{\omega}^{8}t^{6}},  \label{|f00|2}
\end{equation}
where $\tilde{\bar{\omega}}=\sqrt{\bar{\omega}^{2}-\frac{\pi^{2}g^{2}}{4}}$.
In the above expression the approximation $t>>\frac{1}{\bar{\omega}}$ plays
a role only in the two last terms, due to the difficulties to evaluate
exactly the integral in Eq. (\ref{f00}) along the imaginary axis. The first
term comes from the residue at $\Omega=\tilde{\bar{\omega}}+i\frac{\pi g}{2}$
and would be the same if we have done an exact calculation. If we consider
the case of weak coupling, $g<<\bar{\omega}$, we obtain the well known
perturbative exponential decay law for the harmonic oscillator\cite{Beck1}, 
\begin{equation}
|f^{00}(t)|^{2}\approx e^{-\pi gt},  \label{|f00|2'}
\end{equation}
but we emphasize that Eq.(\ref{|f00|2}) is valid for all values of the
coupling constant $g$, even large, it is an expression valid for weak as
well as strong couplings.

- $N=2$

Let us call $\Gamma_{11}^{\mu \nu}$ the configuration in which the dressed
oscillators $q^{\prime}_{\mu}$ and $q^{\prime}_{\nu}$ are at their first
excited level and $\Gamma_{2}^{\mu}$ the configuration in which $%
q^{\prime}_{\mu}$ is at its second excited level. Taking as initial
condition that the dressed mechanical oscillator be at the second excited
level, the time evolution of the state $\Gamma_{2}^{0}$ may be obtained in
an analogous way as in the preceeding case, 
\begin{equation}
\Gamma_{2}^{0}(t)=\sum_{\mu}\left[f^{\mu \mu}(t)\right]^{2}\Gamma_{2}^{\mu} +%
\frac{1}{\sqrt{2}}\sum_{\mu \neq \nu}f^{0\mu}(t)f^{0\nu}(t)\Gamma_{11}^{\mu
\nu},  \label{ortovestidas7}
\end{equation}
where the coefficients $f^{\mu \mu}$ and $f^{0\mu}$ are given by (\ref{fmunu}%
). Then it easy to get the following probabilities:

Probability that the dressed oscillator still be excited at time $t$: 
\begin{equation}
P_{0}(t)=|f^{00}(t)|^{4},  \label{decay0}
\end{equation}
probability that the dressed oscillator have decayed at time $t$ to the
first level by emission of a field quantum: 
\begin{equation}
P_{1}(t)=2|f^{00}(t)|^{2}(1-|f^{00}(t)|^{2})  \label{decay1}
\end{equation}
and probability that the dressed oscillator have decayed at time $t$ to the
ground state: 
\begin{equation}
P_{2}(t)=1-2|f^{00}(t)|^{2}+|f^{00}(t)|^{4}.  \label{decay2}
\end{equation}
Replacing Eq.(\ref{|f00|2}) in the above expressions we get expressions for
the probabilities decays valid for any value of the coupling constant. In
the particular case of weak coupling we obtain the well known perturbative
formulas for the oscillator decay \cite{Beck1}, 
\begin{equation}
P_{0}(t)\approx e^{-2\pi gt},  \label{decay'0}
\end{equation}
\begin{equation}
P_{1}(t)\approx 2e^{-\pi gt}(1-e^{-\pi gt})  \label{decay'1}
\end{equation}
and 
\begin{equation}
P_{2}(t)\approx 1-2e^{-\pi gt}+e^{-2\pi gt}.  \label{decay'2}
\end{equation}

\section{Concluding Remarks}

In this paper we have analysed a symplified version of an
atom-electromagnetic field system and we have tried to give the more exact
and rigorous treatment we could to the problem. We have adopted a general
physicist'
s point of view, in the sense that we have rennounced to approach very
closely to the real behaviour of a complicated non-linear system, to study
instead a simple linear model. As a counterpart, an exact solution has been
possible. Our dressed coordinates give a description of the behaviour of the
system that is exact and valid for weak as well as for strong coupling. If
the coupling between the mechanical oscillator and the field is weak, we
recover the well known behaviour from perturbation theory.

\section{In Memoriam}

This paper evolved from umpublished work we have done and discussions we
have had, with Prof. Guido Beck when two of us (A.P.C.M. and N.P.A.) were
his students at Instituto de Fisica Balseiro in Bariloche (Argentina), in
the late sixties and the early seventies. We dedicate this article to his
memory.

\section{Acknowlegements}

This paper was supported by Conselho Nacional de Desenvolvimento Cientifico
e Tecnologico (CNPq) - Brazil.

\newpage

\begin{figure}

\end{figure}

\end{document}